# Antibonding ground states in semiconductor artificial molecules

M.F. Doty[1*], J. I. Climente[2,3], M. Korkusinski[3], M. Scheibner[1], A.S. Bracker[1], P. Hawrylak[3#] and D. Gammon[1#]

[1]Naval Research Laboratory, Washington, D.C. 20375, USA

[2]CNR-INFM National Center on nanoStructures and bioSystems at Surfaces (S3), Via Campi 213/A, 41100 Modena, Italy

[3]Institute for Microstructural Sciences, National Research Council of Canada, Ottawa, Canada K1A 0R6

**The spin-orbit interaction is a crucial element of many semiconductor spintronic technologies.[1] Here we report the first experimental observation, by magneto-optical spectroscopy, of a remarkable consequence of the spin-orbit interaction for holes confined in the molecular states of coupled quantum dots.[2] As the thickness of the barrier separating two coupled quantum dots is increased, the molecular ground state changes character from a bonding orbital to an antibonding orbital.[3] This result is counterintuitive, and antibonding molecular ground states are never observed in natural diatomic molecules. We explain the origin of the reversal using a four band *k.p* model that has been validated by numerical calculations that account for strain. The discovery of antibonding molecular ground states provides new opportunities for the design of artificially structured materials with complex molecular properties that cannot be achieved in natural systems.**

We investigate single 'artificial diatomic molecules' formed by two vertically stacked self-assembled InAs quantum dots in a GaAs matrix. In general, the dots have different size, composition and strain, and therefore different confined energy levels. As a result, the electron and hole tend to localize in individual dots, as depicted in the left insets of Fig. 1, rather than form delocalized molecular states. Delocalized molecular states are formed when an electric field tunes the relative energies of confined states in the two dots through resonance.[2,4,5] At the resonance, an energy anticrossing occurs, a signature of coherent coupling mediated by particle tunneling between dots.[6,7] The tunnel coupling creates molecular states that are the symmetric and anti-symmetric combinations of the basis states where the particle is in one dot or the other. In analogy to real molecules, we call the symmetric molecular state, which has an enhanced wavefunction amplitude in the barrier, a bonding state. The antisymmetric state has a suppressed amplitude in the barrier and is called the antibonding state. As we will discuss, the bonding/antibonding character of the hole can be identified through the electric field dependence of the g-factor, which is sensitive to the wavefunction in the barrier.

In general, either electron or hole tunneling can be induced[8], but in this work we focus only on hole tunneling, which we will show is especially interesting. Valence holes in semiconductor nanostructures are usually regarded as positively charged particles, much like the conduction electrons, but with heavier effective mass.[9,10] There is, however, an intrinsic difference between conduction electrons and valence holes - the valence states are derived from p-type atomic orbitals of the semiconductor lattice, while electrons are derived from s-type orbitals. Thus, unlike an electron, a hole experiences a strong spin-orbit (SO) interaction.[11] The consequences of the SO interaction for holes appear, for example, in interactions with other holes[12-14] and with the environment.[15,16] Holes in quantum dots can have surprisingly long spin relaxation times[16] and have extremely weak hyperfine interactions with nuclei,[17] which suppresses a primary decoherence mechanism and renders these particles very promising for the development of spin-based quantum information processing.

Another intriguing property of holes has been recently predicted by several atomistic simulations: the molecular ground state for holes in artificial diatomic molecules can have very large antibonding character.[18-21] No such antibonding ground state is observed in natural molecules. In atomic physics the SO interaction has long been known to break parity symmetry and mix the bonding and antibonding character of natural molecules. However, the molecular ground state always remains mostly bonding. For diatomic molecules made of heavy elements, where relativistic corrections are most important, SO interaction contributes up to 10% antibonding character to the otherwise bonding orbitals.[22]

In this Letter we report the first experimental evidence that antibonding molecular ground states can be formed in artificial molecules. We study our artificial quantum dot molecules (QDMs) with photoluminescence spectroscopy. Due to the large inhomogeneous distribution of parameters in ensembles of QDMs, all spectroscopy is performed on single QDMs (see supplementary material). Fig. 1 shows an anticrossing of the neutral exciton in which a single hole tunnels through a thin (2 nm) barrier while the electron remains localized in the bottom dot. At the anticrossing point, the molecular wavefunctions are bonding and antibonding combinations of the individual dot's wavefunctions,[23] as depicted schematically by the right insets in Fig. 1. Intuitively one expects the molecular ground state to have bonding orbital character and the first molecular excited state to have antibonding orbital character, but the orbital character of the molecular states cannot be verified from photoluminescence spectra like that of Fig. 1. Using magneto-photoluminescence measurements, however, we can directly measure the orbital character of the molecular states.[3]

When we apply a magnetic field to InAs/GaAs QDMs there is a large change in the effective g factor at hole tunneling resonances, as described for QDMs with a barrier thickness of 2 nm in Ref. [3]. This effect is shown in Fig. 2e, where we plot the photoluminescence lines of the same QDM shown in Fig. 1, but now in a longitudinal magnetic field of B = 6 T. The resonant change in g factor manifests as a resonant change in Zeeman splitting, which can be seen more clearly in Fig. 2a. The black shading indicates the resonant change in Zeeman splitting for the molecular ground state, while the red shading indicates the resonant change in Zeeman splitting for the first molecular excited state. The resonant changes in Zeeman splitting arise from the contribution of the GaAs barrier.[24]

The resonant amplitude of the wavefunction in the barrier depends on the orbital character. Bonding orbitals have an enhanced amplitude in the barrier, which adds a significant positive component[25,26] to the otherwise negative heavy-hole g factor in InAs dots.[27] As a result, the net Zeeman splitting decreases on resonance. In Fig. 2a, the molecular ground state (black) has a resonant decrease in Zeeman splitting, which indicates that it has the expected bonding orbital character. In contrast, the antibonding wavefunction has a diminished amplitude in the GaAs barrier and consequently a reduced contribution from the barrier and an enhanced Zeeman splitting. In Fig. 2a, the first molecular excited state (red) has a resonant increase in Zeeman splitting, which indicates that it has antibonding orbital character.

In Fig. 2b, 2c and 2d we apply this technique to samples with increasing barrier thickness, *d*. Surprisingly, when *d* = 3 and 4 nm (Fig. 2b and 2c), there is a *reversal* in the nature of the resonance: the molecular ground state (black shading) now shows a resonant *increase* in Zeeman splitting, which conclusively demonstrates that this state now has antibonding character. For the same samples, the first molecular excited state (red shading) now shows a resonant *decrease* in Zeeman splitting indicative of bonding orbital character. The amplitude of the resonant change in Zeeman splitting decreases as the thickness of the barrier increases, and is below our noise level for d = 6 nm (Fig. 2d). This results from the reduction of the amplitude of the wavefunction in the barrier with increasing barrier thickness.

The counterintuitive reversal of the orbital character of the molecular ground state is a consequence of the SO interaction. With SO interaction, the hole's atomic orbital and spin degrees of freedom couple to give a total (Bloch) angular momentum *J=3/2*, where $J_z = \pm 3/2$ projections correspond to heavy holes and $J_z = \pm 1/2$ to light holes. The light hole states are shifted up in energy by confinement and strain. As a result it is often useful to represent the two low-energy (heavy-hole) states as pseudo-spin ½ particles (⇓,⇑). When the heavy-light hole mixing is included, low-energy hole states can be described within the Luttinger-Kohn $k{\cdot}p$ Hamiltonian formalism[11] as four component Luttinger spinors.[12] Each spinor is an admixture of all four projections of $J_z$, with the heavy hole typically the dominant component. However, as we show here, minor components of no more than 5% are sufficient to substantially alter the character of the molecular orbitals in QDMs.

The influence of the minor components is apparent from Fig. 3. In Fig. 3a, we plot the energies of the molecular ground and first excited states calculated using a simple one-band effective mass model, which neglects SO interactions. As expected, the energy separation of the bonding (solid black line) and antibonding (dashed red line) states decreases as a function of increasing barrier thickness and the bonding orbital remains the molecular ground state. In Fig. 3b we show the energies of the bonding (solid blue line) and antibonding (dashed red line) states calculated using a four-band $k{\cdot}p$ model that includes the SO interaction. At $d$~1.75 nm the energies of the bonding and antibonding states cross and the antibonding state becomes the molecular ground state. From our model, we estimate the antibonding character of the ground state spinor for large barrier thicknesses to be as large as 95%, many times larger than in known atomic systems (see supplementary material for details).

The formation of molecular orbitals at an anticrossing (for example, Fig. 1) is described by a simple Hamiltonian using an atomic-like basis with a hole either in one dot or the other:

$$\begin{pmatrix} E_0 & -t \\ -t & E_0 - f + f_0 \end{pmatrix} \qquad (1)$$

Here $E_0$ is the energy of the localized hole states at resonance, $t$ is the tunneling rate, and $f = e\tilde{d}F$ is the Stark energy due to the electric field $F$. The energy of the bonding and antibonding molecular orbitals are given by the eigenvalues of Eq. 1. When the electric field is tuned to resonance ($f = f_0$), the energies are $E_b$=$E_0$ - $t$ and $E_{ab}$=$E_0$ + $t$.

The tunneling rate in Eqn. 1 is determined by the energy splitting between the antibonding and bonding states: $2t = E_{ab} - E_b$ . The values for $t$ corresponding to the two cases of Fig. 3a and 3b are plotted in Fig. 3c by $t_0$ and $t$, respectively. In the absence of SO interaction, the tunneling rate is determined by the overlap between the hole orbitals of the individual dots ($t_0$), which decreases exponentially with increasing barrier thickness at a rate dependent on the heavy-hole mass. When the SO interaction is included, there is a correction to the tunneling rate, $t = t_0 - t_{so}$. This $t_{so}$ term arises from the small contribution of the light-hole component of the spinor. The light-hole component has approximate parity along $z$ opposite to that of the heavy-hole component (see supplementary material). The light-hole component therefore adds a small antibonding (bonding) component to the bonding (antibonding) state determined by the dominant heavy-hole component, as shown schematically in Fig. 3d. The addition of this antibonding component increases the energy of the bonding state (and vice versa for the antibonding state). As the barrier thickness increases, $t_{so}$ does not decrease as fast as $t_0$, in part because of its light-hole origin. For thin barriers, the $t_{so}$ correction is small compared to the large $t_0$, and $t$ remains positive. However, for thicker barriers $t_0$ decreases and becomes comparable to $t_{so}$, first leading to $t = 0$ and then to negative tunneling rates (when $t_0 < t_{so}$). The negative tunneling rate corresponds to the antibonding molecular ground state.[18-21]

We have verified this simple four-band $k{\cdot}p$ picture against an atomistic multi-million atom calculation of the hole levels of a QDM described by the sp$^3$d$^5$s* tight-binding model.[28] This approach accounts for strain and changes to the underlying crystal lattice on the atomistic level (see supplementary material for further details). The results of this calculation are shown by the solid blue points in Fig. 3a, which qualitatively match the results of the $k{\cdot}p$ calculation. Thus, the $k{\cdot}p$ approach is sufficient to capture the essential physics of the system.

We now return to the reversal of the Zeeman splitting resonance shown in Fig. 2 and present a quantitative analysis of the tunneling rate. The resonant change in Zeeman splitting enters into Eq. 1 as a spin-dependent contribution to the tunneling rate ($-t \Rightarrow -t \pm h'$). Essentially, the potential barrier height is increased or decreased by the barrier contribution to the Zeeman energy, $h'$. The normal Zeeman energy for the isolated dots, $h_0 = (g_e + g_h)\mu_B B/2$, appears on the diagonal. Including these terms, we obtain two Hamiltonians for the two spin configurations $\downarrow\Uparrow$ and $\uparrow\Downarrow$ (where $(\uparrow,\downarrow)$ and $(\Downarrow,\Uparrow)$ are the electron spin and hole spinor projections)

$$\begin{pmatrix} E_0 - h_0 & -t - h' \\ -t - h' & E_0 - h_0 - f + f_0 \end{pmatrix} \text{ and } \begin{pmatrix} E_0 + h_0 & -t + h' \\ -t + h' & E_0 + h_0 - f + f_0 \end{pmatrix} \qquad (2)$$

The Zeeman energy terms have opposite signs: $h' > 0$ for a GaAs barrier and $h_0 < 0$ because both the electron and hole g factors ($g_e, g_h$) are negative for InAs dots.[27] The eigenvalues of Eq. 2 give the enhanced and suppressed energy anticrossings for the two spin states. To obtain an expression for the Zeeman splitting of the ground and excited states, we take the difference between the eigenenergies in Eq. 2.

$$\Delta_{G(E)} = 2h_0 \pm \frac{2th'}{\sqrt{(f - f_0)^2 + 4t^2}} \qquad (3)$$

Because the signs of $h'$ and $h_0$ are fixed, the sign of $t$ determines whether the Zeeman splitting of the ground state is enhanced or suppressed at the resonance. In the 4 nm case (Fig. 2c), the Zeeman splitting of the molecular ground state ($\Delta_G$) increases at the anticrossing point. $t$ must therefore be negative, which confirms that the molecular ground state has antibonding character. In contrast, the Zeeman splitting of the upper level ($\Delta_E$) comes from the minus sign in Eq. 3, so a negative $t$ leads to a decrease in the splitting at resonance, as observed. This excited molecular orbital must therefore have bonding character. The solid curves in Fig. 2a, 2b and 2c are calculated by fits to Eq. 3. From these fits, we find that $t$ is positive for the QDM with $d$ = 2 nm (Fig. 2a) and negative for QDMs with $d$ = 3 and 4 nm (Fig. 2b and 2c), which confirms the reversal in the sign of $t$ predicted by the model.

Our theoretical model predicts that the antibonding state becomes the molecular ground state for a barrier thickness $d \geq 1.75$ nm. Experimentally, we find that all examples (7) in the sample with $d = 2$ nm show that the molecular ground state has bonding orbital character; in the sample with $d = 4$ nm, all examples (3) show that the ground state has antibonding character.  For the intermediate case ($d = 3$ nm), we find examples of both types of behavior, which indicates that the reversal of orbital character occurs near $d = 3$ nm. The co-existence of both behaviors at $d = 3$ nm most likely arises from the inhomogeneous distribution of specific dot parameters. The discrepancy between the predicted and observed barrier thickness at which the orbital character switches is likely due to details of dot structure, composition and alignment that are not accounted for in the theoretical models.

The orbital character of the molecular ground state can also be altered by the addition of more holes. In Fig. 4a we schematically depict the filling of the molecular orbitals when the bonding state is the lowest energy single particle state (i.e. for d= 2 nm).  A second hole can go into the same orbital, and so the lowest energy two-body state is a spin singlet with bonding character.  Because it is a spin singlet there is no magnetic field spin splitting.  The lowest energy three-hole state must have an unpaired hole in the antibonding orbital.  Because of the unpaired hole, this state should have a magnetic field splitting like that of the one-hole state, but the Zeeman splitting should increase on resonance because the unpaired hole is in the antibonding orbital. This is exactly what we observe, as shown in Fig. 4c. The one-hole and three-hole states of the same QDM (with d = 2nm), have opposite behavior, indicating that they have different orbital character. The three hole state is observed as the initial state of the doubly charged exciton (three holes and one electron) (see supplementary material).

In this letter we have presented experimental evidence and an intuitive explanation of a remarkable consequence of the SO interaction, which is a fundamental element of semiconductor spintronics.[1] The SO-induced reversal of the orbital character of the molecular ground state for holes in artificial diatomic molecules provides a powerful tool for the design of spin manipulation protocols[29] and artificial molecules with complex properties not achievable in natural molecules.

We acknowledge support from NSA/ARO, ONR, CIAR, QuantumWorks, FIRB-MIUR Italy-Canada RBIN06JB4C, MEIF-CT-2006-023797 (J.I.C.) and Institute for Microstructural Sciences(J.I.C.).

* Matt Doty is now at University of Delaware (doty@udel.edu)

# Correspondence and requests for materials should be addressed to: Dan Gammon (gammon@nrl.navy.mil) and Pawel Hawrylak (pawel.hawrylak@nrc-cnrc.gc.ca)

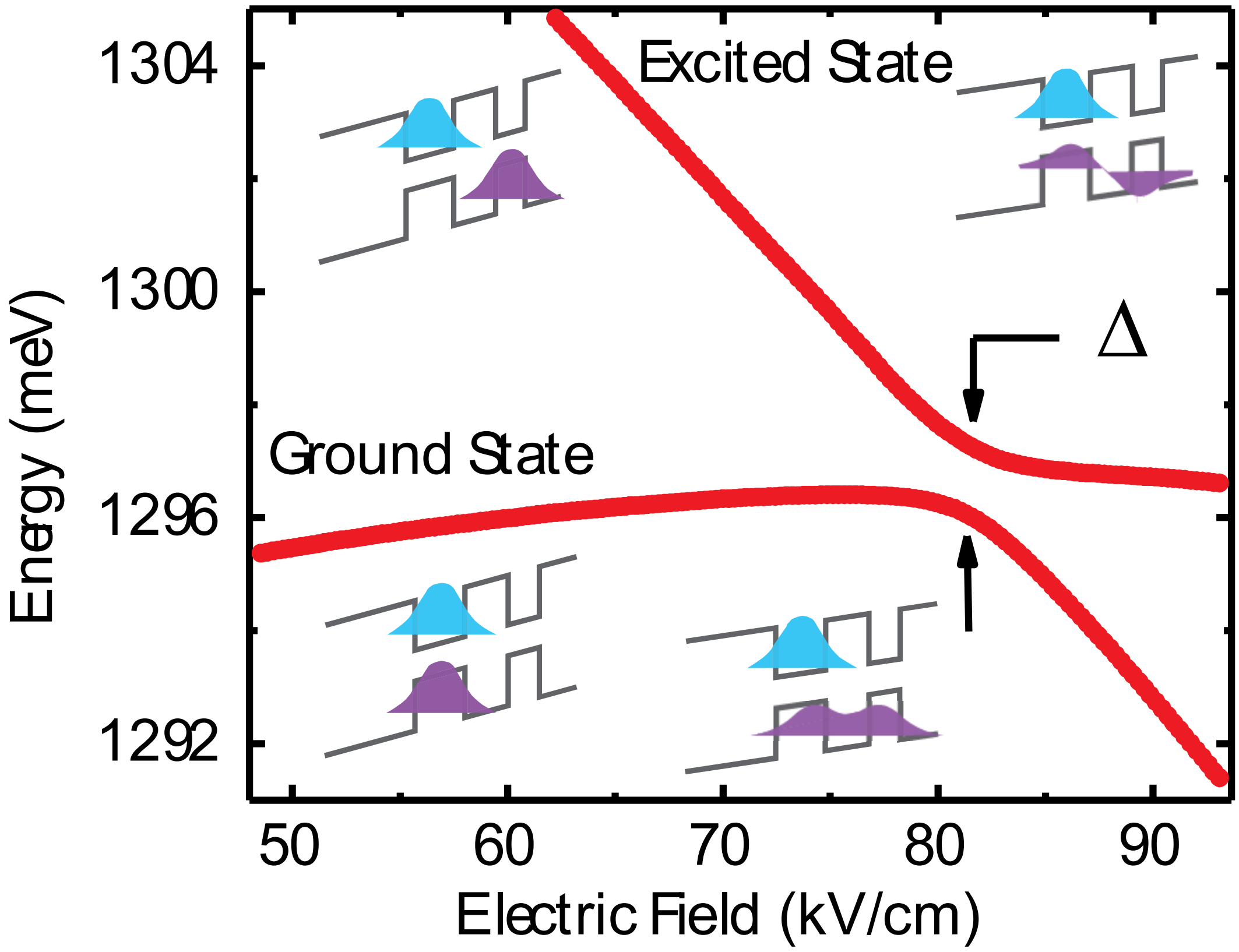

Fig. 1. Electric-field induced anticrossing of the neutral exciton (single electron hole pair) at zero magnetic field for a sample with 2 nm barrier. Δ indicates the anticrossing energy gap ( $\Delta = |E_{ab} - E_b| = 2|t|$ ). Insets: If the hole energy levels are out of resonance (left) the hole is localized in one of the individual dots. When the hole levels are tuned into resonance by the applied electric field, coherent tunneling leads to the formation of bonding (bottom right) and antibonding (upper right) molecular wavefunctions.

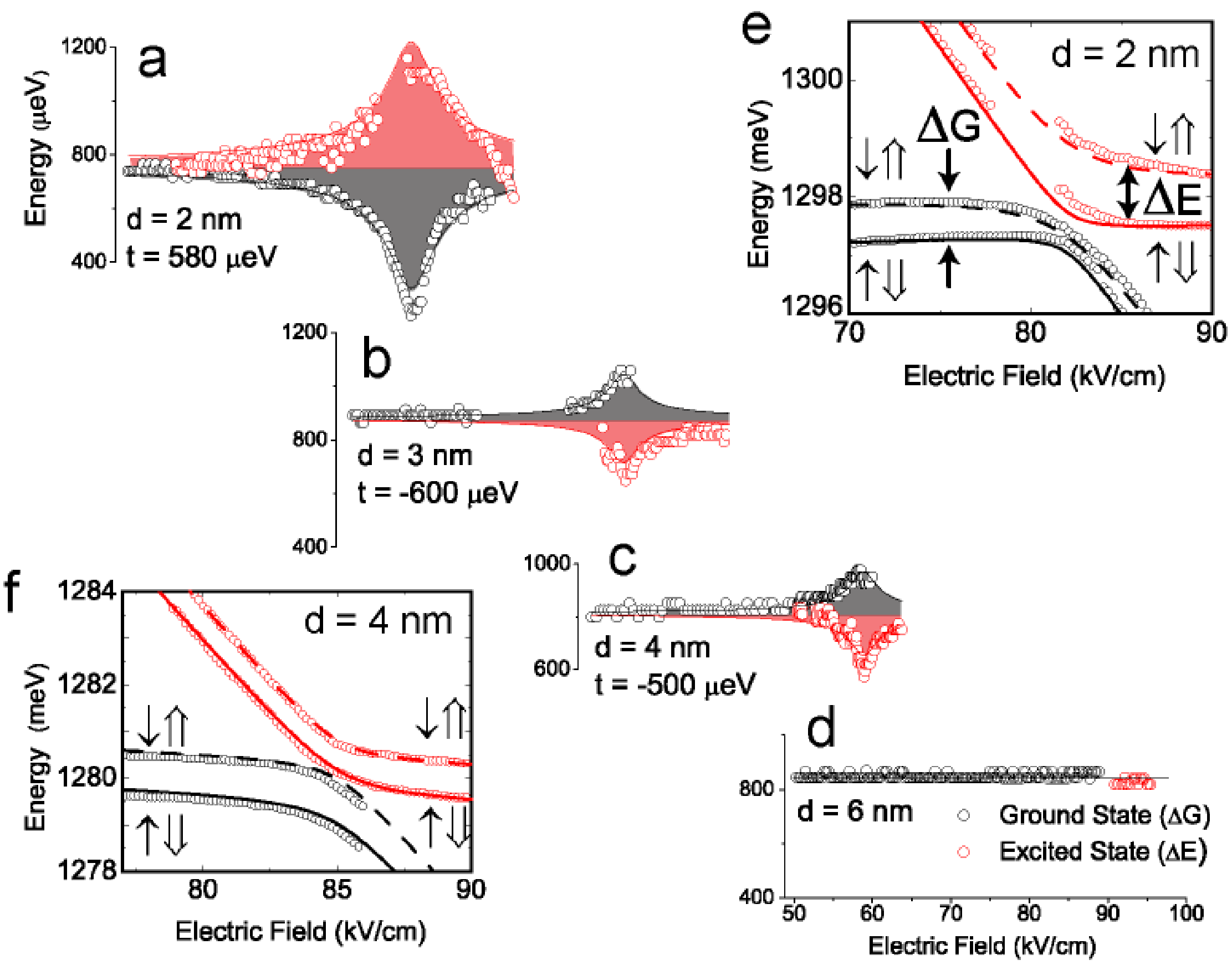

Fig. 2. Zeeman energy splitting as a function of applied electric field for B=6 T. The QDM has a barrier thickness of (a) 2, (b) 3, (c) 4, and (d) 6 nm. The solid curves are calculated with Eq. 3 using ($h_0$, $h'$) = (-0.368, 0.229), (-0.446, 0.082), (-0.404, 0.076) meV for 2, 3, and 4nm, respectively. (e, f) Energies of the neutral exciton photoluminescence lines as a function of applied electric field for the 2 nm barrier (e) and 4 nm barrier (f). Solid and dashed lines indicating the two separate spin configurations (labeled) are calculated using Eq. 2 (see supplementary material for details).

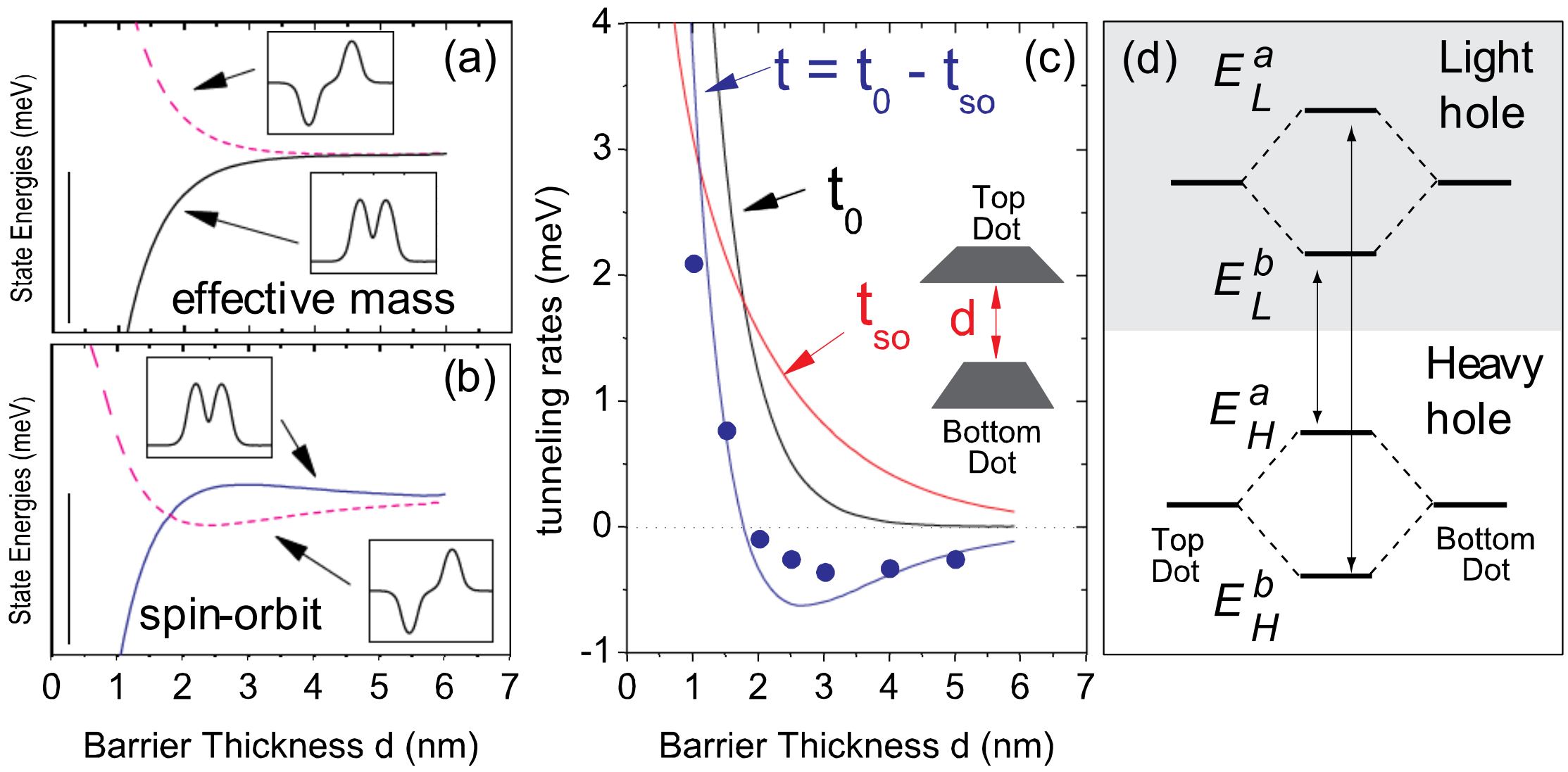

Fig.3. (a, b) Energy of the ground and first excited molecular states for asymmetric QDMs subject to resonant electric field as a function of barrier thickness calculated using the effective mass (a) and $k{\cdot}p$ theory including spin-orbit interactions (b). Scale bars are 5 meV. The insets illustrate the profile of the dominant hole spinor component along the molecular coupling direction. (c) Tunneling rates of a single hole versus barrier thickness. The tunneling rates $t$, $t_0$, and $t_{so}$ are calculated using single-band effective mass (black line) and $k{\cdot}p$ theory (solid blue line). Blue points are calculated with an atomistic simulation that includes strain. The inset provides a schematic of the coupled quantum dots. (d) Schematic depiction of the bonding ($E^b$) and antibonding ($E^a$) molecular states for the heavy ($E_H$) and light ($E_L$) holes. The spin-orbit interaction mixes $E_H^b$ with $E_L^a$ and $E_H^a$ with $E_L^b$.

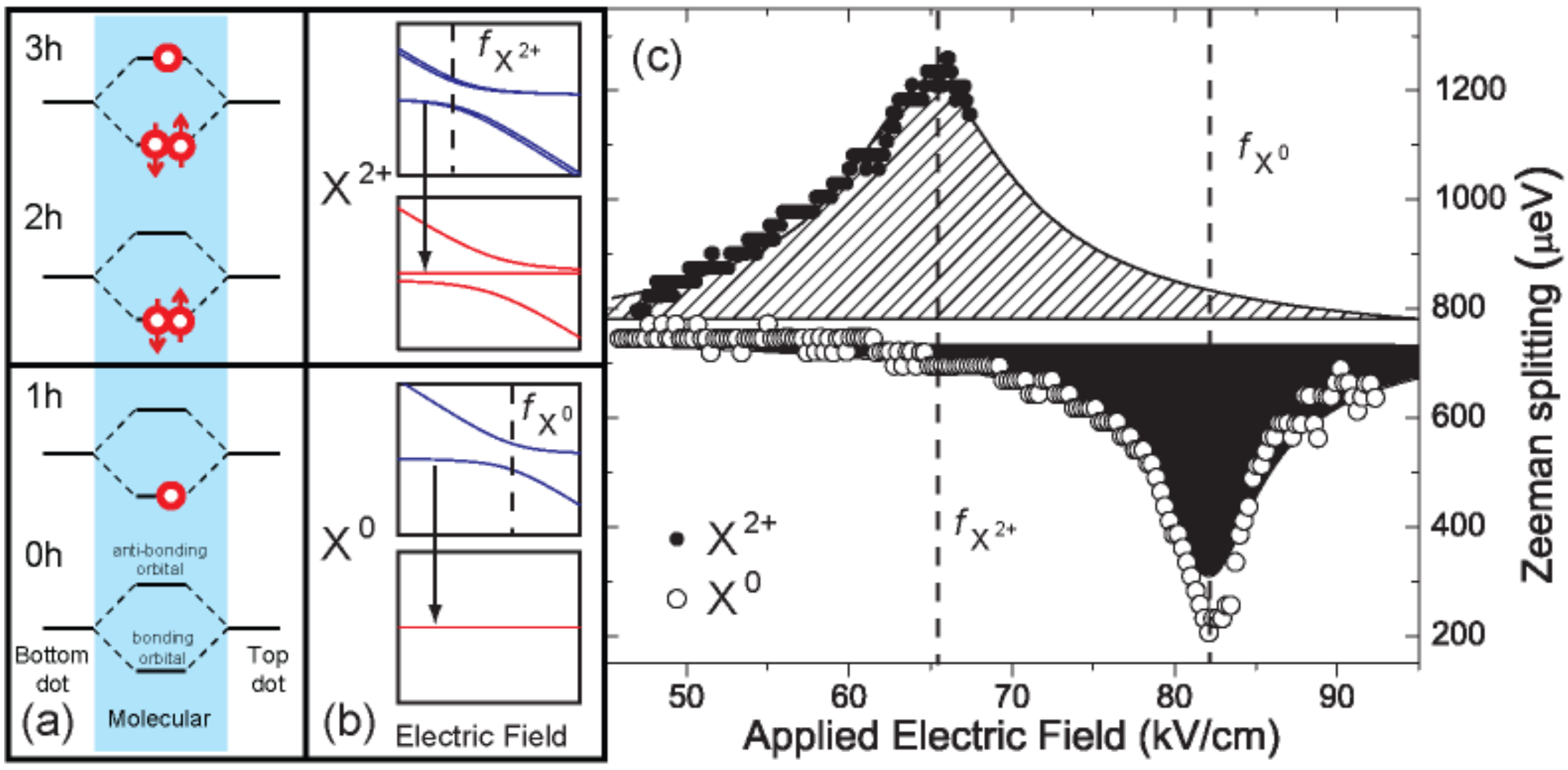

Fig.4. (a) Molecular ground states for higher charge configurations are determined by sequentially filling the molecular orbitals with additional holes. (b) Calculated energy levels for the $X^{2+}$ transition, which starts in a state with 3 holes (and 1 electron) and ends in a state with 2 holes, and the $X^0$ transition, which starts in a state with 1 hole (and 1 electron) and ends in a state with 0 holes. (c) Electric field dependence of the Zeeman splitting for the molecular ground states of the $X^0$ and $X^{2+}$ transitions in a QDM with 2nm barrier at B = 6 T. $X^0$ data is the same as in Fig. 2. The $X^0$ and $X^{2+}$ resonances peak at two different values of the electric field $f_{X^0}$ and $f_{X^{2+}}$ because of different Coulomb interactions.

**Antibonding ground states in semiconductor artificial molecules**

# Supplementary Material

M.F. Doty, J. I. Climente, M. Korkusinski, M. Scheibner, A.S. Bracker, P. Hawrylak and D. Gammon

## 1. Sample growth and measurement technique

The quantum dot molecules (QDMs) were fabricated by the successive molecular beam epitaxial growth of two closely spaced layers of self assembled InAs/GaAs quantum dots (QDs). Strain causes the dots of the second layer to nucleate preferably on top of dots in the first layer, thereby forming QDMs. The vertical height, z, of the QDs was controlled by the application of an indium flush technique.[30] Here the InAs islands are partially covered with a GaAs layer of thickness z. Then the deposition of material is interrupted while the temperature is raised, causing the uncovered InAs to redistribute. After that the deposition of GaAs is continued at the regular growth temperature. The vertical height of the QDs determines the ground state transition energy and was chosen to be 2.5 nm in order to achieve photoluminescence energies between 930 and 1000 nm. In order to tune hole levels into resonance by applying an electric field in the growth direction, the QDMs were embedded in an n-type Schottky diode structure.

Individual QDMs were optically excited and detected through an aluminum shadow mask with 1 μm diameter apertures. Excitation was performed with a continuous wave titanium-sapphire laser tuned to wavelengths between 895 and 915 nm, i.e. energetically well below the wetting layer emission at about 870 nm. The sample was placed in the bore of split coil superconducting magnet and positioned under the focus of a 0.45 NA lens using low-temperature non-magnetic translation stages manufactured by Attocube. The photoluminescence signal was dispersed with a 0.75 m monochromator equipped with an 1100 mm-1 line grating, and was detected by a liquid nitrogen cooled charged coupled device (CCD) camera. The overall spectral resolution of this system was about 70 μeV.

Spectral maps of these QDMs are obtained by assembling line spectra acquired at sequential values of the applied electric field. These spectral maps contain many charge states whose lines can cross and overlap. Neutral exciton lines can be identified from the characteristic negative trion charging energy, the lack of x-patterns [4], and characteristic spin fine structure.[31] The optical intensities of the lines can vary by several orders of magnitude as the optical recombination changes character from direct (electron and hole in the same dot) to indirect (electron and hole in different dots). To aid the reader, in the main text the energies of the neutral exciton photoluminescence lines have been extracted and plotted.

Solid lines in Figure 2a-d are calculated using Equation 3. The values of $t$, $\tilde{d}$ and $f_0$ are determined from the experimental spectra taken at zero magnetic field. Because of alloying, the *effective* barrier thickness, $\tilde{d}$, is slightly different from the nominal (growth controlled) dot separation, $d$. $\tilde{d}$ is determined by the measured slope of the indirect photoluminescence line resulting from recombination of an electron in the bottom dot with a hole in the top dot. $\tilde{d}$ enters into the equations as $f = e\tilde{d}F$ , which describes the Stark shift of the indirect (spatially separated) exciton. The energies of the indirect and direct excitons are degenerate at the resonant electric field, $f_0$ , which is measured from the maximum of the anticrossing gap. The value of $h_0$ is determined by the asymptotic value of the Zeeman splitting away from the anticrossing. The value of $h'$ is determined by fitting Equation 3 to the experimental data. Solid lines in Figures 3b and 3c are calculated with Equation 2 using the values described above, including the value of $h'$ determined by the fit to Equation 3. $E_0$ is determined by the absolute energy of the experimental data. To account for the stark shift of the exciton confined in the bottom dot, a linear dependence of the energy on electric field is added to the (1,1) element of the matrix. $\tilde{d}$ and this additional single dot Stark shift slope are determined by the experimental data.

## 2. k·p model

We model the quantum dot molecule as two vertically coupled quantum disks with circular symmetry. The confining potential is $\hat{V}(\rho,z) = \hat{V}(\rho) + \hat{V}(z) + E_z \cdot z$ . Here $\hat{V}(\rho)$ is an infinite barrier in the radial direction, $\hat{V}(z)$ is a double square well potential, whose height is the band-offset between the dot and barrier materials, $V_c$, and $E_z$ is an electric field applied along the vertical direction.

The 4-band Luttinger-Kohn Hamiltonian[11], which considers the coupling between heavy hole ($J = 3/2, J_z = \pm 3/2$) and light hole ($J = 3/2, J_z = \pm 1/2$) subbands, is spanned in the basis $J_z$=+3/2,-1/2,+1/2,-3/2, so that it reads:

$$\hat{H}_{LK} = \begin{pmatrix} \hat{P}_+ & \hat{R} & -\hat{S} & 0 \\ \hat{R}^* & \hat{P}_- & 0 & \hat{S} \\ -\hat{S}^* & 0 & \hat{P}_- & \hat{R} \\ 0 & \hat{S}^* & \hat{R}^* & \hat{P}_+ \end{pmatrix} + \hat{V}(\rho,z)\,I \qquad (1)$$

where $I$ is the identity matrix. The operators in the above expression are given by:

$$\hat{P}_+ = \frac{\eta^2}{2}\left[(\gamma_1+\gamma_2)\,\hat{p}_\perp^2 + (\gamma_1 - 2\gamma_2)\,\hat{p}_z^2\right], \qquad (2)$$

$$\hat{P}_- = \frac{\eta^2}{2}\left[(\gamma_1-\gamma_2)\,\hat{p}_\perp^2 + (\gamma_1 + 2\gamma_2)\,\hat{p}_z^2\right], \qquad (3)$$

$$\hat{R} = \frac{\eta^2}{2}(-\sqrt{3})\,\gamma_2\,\hat{p}_-^2, \qquad (4)$$

$$\hat{S} = \frac{\eta^2}{2}(2\sqrt{3})\,\gamma_3\,\hat{p}_-\hat{p}_z, \qquad (5)$$

with the momentum operators $\hat{p}_z = -i\nabla_z$, $\hat{p}_\pm = -i(\nabla_x \pm i\nabla_y)$ and $\hat{p}_\perp^2 = \hat{p}_x^2 + \hat{p}_y^2$.

The hole state with a total angular momentum $F_z$ and chirality symmetry ν can be written as a four-component Luttinger spinor:

$$
\left|F_z,\nu\right\rangle = \sum_{n,l} \begin{pmatrix} A^{Fz,\nu}_{+3/2,n,l}\, f_{F_z-3/2,n}(\rho,z)\,\xi_l(z)\left|J_z = +\frac{3}{2}\right\rangle \\ A^{Fz,\nu}_{-1/2,n,l}\, f_{F_z+1/2,n}(\rho,z)\,\xi_l(z)\left|J_z = -\frac{1}{2}\right\rangle \\ A^{Fz,\nu}_{+1/2,n,l}\, f_{F_z-1/2,n}(\rho,z)\,\xi_l(z)\left|J_z = +\frac{1}{2}\right\rangle \\ A^{Fz,\nu}_{-3/2,n,l}\, f_{F_z+3/2,n}(\rho,z)\,\xi_l(z)\left|J_z = -\frac{3}{2}\right\rangle \end{pmatrix} \tag{6}
$$

where $\left|J_z\right\rangle$ is the Bloch function and $f_{m_z,n}(\rho,\theta)$ is the in-plane envelope part ($m_z$=$F_z$-$J_z$),

$$
f_{m_z,n}(\rho,\theta) = \frac{e^{i\,m_z\,\theta}}{\sqrt{2\pi}}\,\frac{\sqrt{2}}{R}\,\frac{J_{m_z}(k_n^{m_z}\rho)}{\left|J_{m_z+1}(k_n^{m_z}R)\right|}. \tag{7}
$$

$J_{m_z}(k_n^{m_z}\rho)$ is the Bessel function of order $m_z$ and radial quantum number $n$. $k_n^{m_z}$ represents the hole wave vector, defined in terms of the Bessel function roots ($\alpha_n^{m_z}$) as $k_n^{m_z} = \alpha_n^{m_z}/R$. The vertical components in Eq. (6) are trigonometric functions, $\xi_l(z) = \sqrt{\frac{2}{W}}\cos\left(\frac{l\pi z}{W}\right)$ for $l$ odd, and $\xi_l(z) = \sqrt{\frac{2}{W}}\sin\left(\frac{l\pi z}{W}\right)$ for $l$ even, where $W$ represents the size of the computational box along $z$, which is extended well beyond the two dots and the intermediate barrier.

Hole states are calculated by exact diagonalization of $\hat{H}_{LK}$, on a basis with 6 radial states ($n$=0-5), and 60 harmonics in the vertical direction ($l$=1-60). Single-band (effective mass model) hole states are obtained in the same way, but setting the off-diagonal terms of $\hat{H}_{LK}$ to zero.

For Figure 3 of the paper, we have considered InGaAs/GaAs quantum dot molecules. For simplicity, we have taken InGaAs effective mass inside and outside the dots. The Luttinger parameters are $\gamma_1 = 11.01$, $\gamma_2 = 4.18$, and $\gamma_3 = 4.84$, and the valence band-offset $V_c$=380 meV.[32] The disks have 15 nm radius and the height of the bottom (top) disk is 2.6 (2.4) nm. The anticrossing energy is the minimum energy splitting between lower and upper energy levels as a function of the applied electric field. Further details about the model are given in Ref. [33].

As mentioned in the paper, one can estimate the tunneling matrix elements $t$ from the bonding-antibonding energy splitting, $\Delta = 2|t|$. We define $t_0$ as the tunneling element inferred from the effective mass model, and $t$ as the tunneling element inferred from the four-band Hamiltonian. The effect of the spin-orbit induced band-coupling is then defined as $t_{so} = t - t_0$. For the lowest bonding-antibonding pair of states, $t_{so}$ is negative, hence the negative sign we use in the main text ($t = t_0 - t_{so}$).

To estimate the bonding/antibonding character of the hole wavefunction, we compare the coefficients of the spinor components in Eq.(6). For example, for a chirality up state (ν=↑), the Jz=+3/2 and Jz=-1/2 components are bonding, while the Jz=+1/2 and Jz=-3/2 are antibonding (the opposite holds for chirality down)[33]. For d<2 nm, the ground state is |Fz=3/2, ν=↑>. If we compare the weight of the Jz=+3/2,-1/2 and Jz=+1/2,-3/2 components, we conclude that the ground state is about 95% bonding. Likewise, the first excited state |Fz=3/2, ν=↓> is 95% antibonding. Thus, the spin-orbit induced correction to the pure parity of the wavefunctions is moderate (≤5%). However, this correction has an important effect on the energy with increasing interdot barrier because it is associated with light holes, whose tunneling rates remain significant when heavy hole rates are already small. Thus, at d>2 nm the spin-orbit correction suffices to induce a state reversal and |Fz=3/2, ν=↓> becomes the (mostly antibonding) ground state.

## 3. Atomistic tight-binding model

The atomistic simulations of the electronic structure of the QDM were performed in the frame of the atomistic $sp^3d^5s^*$ tight-binding approach in the NEMO3D implementation.[28,34] The simulation consists of two steps. In the first step we account for the presence of strain caused by the mismatch of the lattice constants of dot and barrier materials by writing the total elastic energy $E_{TOT}$ of the system as a sum of bond-stretching and bond-bending terms for each atomic bond [35-37], and we adjust all atomic positions so as to minimize $E_{TOT}$. Because the resulting displacement field is of a long-range character, the computational domain used in this part of the simulation must be much larger than the QDM. For our dots with a diameter of 15 nm and height of 2.5 nm, positioned on 0.5-nm-thick wetting layers, and separated by a barrier of up to 6 nm we typically employ a domain containing about 25 million atoms.

The equilibrium positions of atoms are further used to compute the electron and hole energies and wavefunctions in the tight-binding approach, in which the single-particle Hamiltonian is written in the form [28,34]:

$$H_{TB} = \sum_{R\alpha} \left( \varepsilon_{R\alpha} + \varepsilon_F \right) c^+_{R\alpha} c_{R\alpha} + \sum_{R\alpha} \sum_{\Delta R\beta}^{n.n} t_{R\alpha,\Delta R\beta} c^+_{R\alpha} c_{\Delta R\beta}, \tag{8}$$

where $c^+_{R\alpha}$ ($c_{R\alpha}$) is the creation (annihilation) of a particle on orbital α of atom *R*, and *ΔR* runs through the nearest neighbors of atom *R*. In our simulation we place 10 spin-degenerate orbitals on each atom. The matrix elements $\varepsilon$ and $t$ are material-specific parameters, modified appropriately by the atomic displacements. The electric field is incorporated into the diagonal elements in the form $\varepsilon_F = eFR$. The above Hamiltonian is written in a form of a matrix of order 20*N*, with *N* being the number of atoms in the computational domain, and the energies and wavefunctions of the electron and the hole are obtained by diagonalizing this matrix numerically. Since we seek only the states confined in the QDM, the computational domain in this part of the calculation is much smaller than that used for the computation of strain (typically we take about 3 million atoms).

## 3. $X^{2+}$

In Fig. 4 we present the resonant change in Zeeman splitting for the $X^{2+}$ transition. The initial state, which is populated by one electron and three holes, has a relatively simple anticrossing similar to the neutral exciton. The only significant difference is that the indirect line (with non-zero slope) in the upper panel of Fig. 4b is split into a doublet. The doublet arises from the bright and dark spin configurations of the electron and single hole confined in the bottom dot, but both spin configurations are optically allowed due to the presence of holes in the top dot. The final state has only two holes and both singlet and triplet spin configurations are possible. The Zeeman splitting plotted in Fig. 4c is for the transition that begins in the low-energy molecular orbital of the $X^{2+}$ state and ends in the triplet state of the final holes with total angular momentum zero. This final state has two holes with antiparallel spins and thus no Zeeman splitting of any kind. For easy comparison, the energy levels of the $X^{2+}$and $h^{2+}$ states are plotted after subtracting a linear dependence of all state energies on applied electric field, which comes from the presence of additional holes in the top dot. A comprehensive description of the $X^{2+}$ state will be presented elsewhere.[38]